\documentclass{kluwer}    
\usepackage{graphicx}

\begin{document}

\begin{article}
\begin{opening}
\title{Dust penetrated morphology in the high redshift Universe}
\author{D.L. \surname{Block$^1$}, I. \surname{Puerari$^2$},
M. \surname{Takamiya$^3$}, R. \surname{Abraham$^4$},
A. \surname{Stockton$^5$}, I. \surname{Robson$^6$},
W. \surname{Holland$^6$}}
\institute{1. University of Witwatersrand, South Africa;
           2. INAOE, Mexico; 3. Gemini Observatory, USA;
           4. University of Toronto, Canada;
           5. University of Hawaii, USA;
           6. Joint Astronomy Center, USA}


\begin{abstract}
Images from the Hubble Deep Field (HDF) North and South show
a large percentage of dusty, high redshift galaxies whose 
appearance
falls
outside traditional classification systems. The nature of these 
objects
is not yet fully understood. Since the HDF preferentially samples
restframe
UV light, HDF morphologies are not dust or `mask'
penetrated.
The appearance of high redshift galaxies at near-infrared
restframes remains a challenge for the New Millennium.
The Next Generation Space Telescope
(NGST) could routinely provide us with such images. In this
contribution,
we quantitatively determine the dust-penetrated structures of high
redshift galaxies such as NGC 922 in their near-infrared restframes.
We show that such optically peculiar objects may readily
be classified using
the dust penetrated z$\sim$0 templates of 
\citeauthor{blockpuerari99}
\shortcite{blockpuerari99} and \citeauthor{butablock01}
\shortcite{butablock01}.
\end{abstract}

\end{opening}

\section{Introduction}

Several studies have recently been conducted to elucidate
the morphology of high redshift objects (e.g., Glazebrook et al.
\citeyear{glazebrooketal95}; Driver et al. \citeyear{driveretal98};
Abraham et al. \citeyear{abrahametal96a};
Abraham et al. \citeyear{abrahametal96b}; Bouwens et al.
\citeyear{bouwensetal98}; Giavalisco et al. 
\citeyear{giavaliscoetal96}).

Near-infrared studies ($\sim$ 2.1 $\mu m$) probe underlying
stellar masses far more accurately than do optical morphologies. In
order to
`penetrate the mask' and
effectively delineate the stellar mass distribution of the older
population, optical detectors do not suffice.

To dust-penetrate galaxies at high redshifts in their K$'$
restframes
with the NGST necessitates the design of detectors sensitive to
light whose
wavelengths are longward of K$'$ (2.1 $\mu m$). A preliminary
description of the NGST facilities are given in Gillett and Mountain
\shortcite{gilletmountain98}. Bally and Morse 
\shortcite{ballymorse99}
describe a detector which will operate in the wavelength range 0.6 
to
5.3 $\mu m$. This detector can observe K$'$ restframe light of 
objects
out to redshifts of $z\approx 1.5$.

\section{Analysis}

The images analysed here were generated following the simulation
methodology of Takamiya \shortcite{takamiya99}. The 
bidimensional
Fourier method was used to secure information about the
morphology (see, e.g., Puerari and Dottori 
\citeyear{pueraridottori92};
Block and Puerari \citeyear{blockpuerari99}, and references 
therein).
The simulations recreate a given image when moved out to higher
redshifts, always in a pre-selected restframe. In this contribution,
the images of NGC 922 are simulated at redshifts $z=0.7$ and
$z=1.2$ in the K$'$ restframe. The original image of NGC 922
was obtained at the NASA Infrared Telescope Facility with NSFCam
(more details can be found in Block et al. \citeyear{blocketal01}).

The bidimensional Fourier analysis shows that the near-infrared
morphology of NGC 922 remains the same when the galaxy is 
moved out
to redshifts $z=0.7$ and $z=1.2$ (see Fig. 1). The
analysis also confirms that only two low-order $m$ components 
($m=1, 2$)
are sufficient to describe the dust-penetrated morphology of NGC 
922
(see Fig. 2).

\begin{figure}
\centerline{\includegraphics[width=35pc]{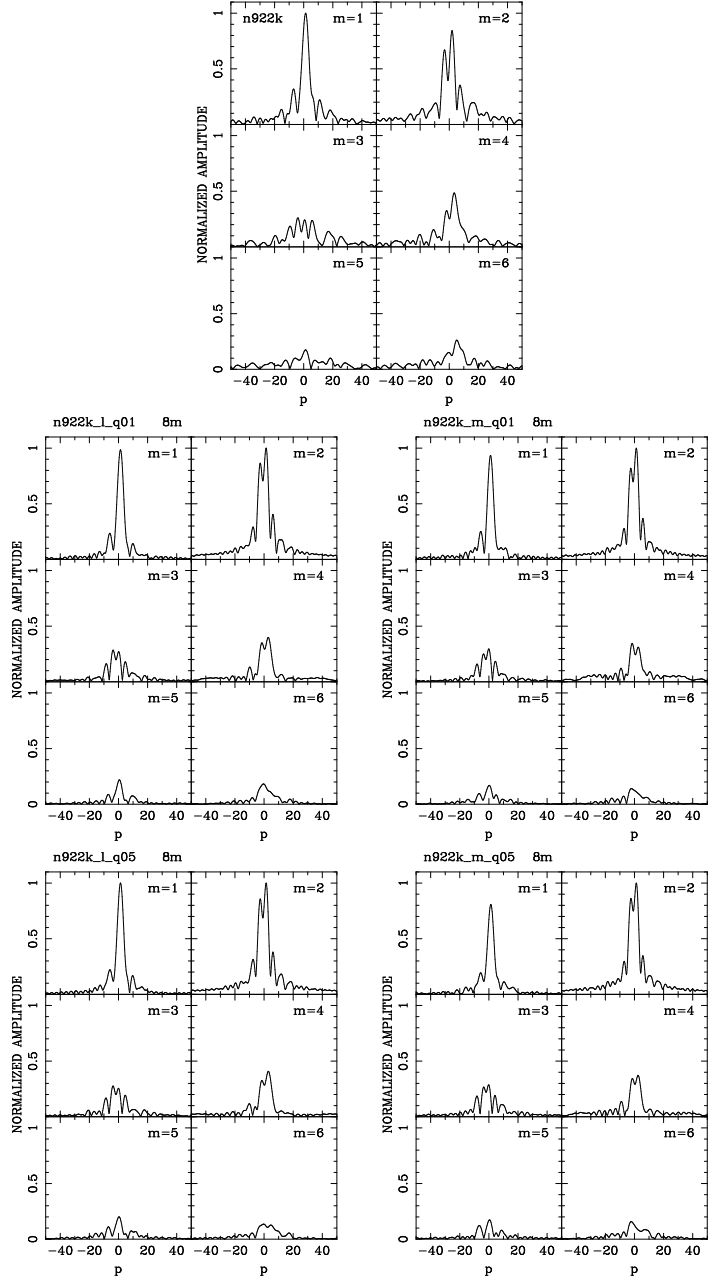}}
\vskip-10pt
\caption[]{The Fourier spectra of the K$'$ image of NGC 922 at its
original distance (top). The middle panels show the Fourier
spectra
when NGC 922 is moved to redshifts $z=0.7$ ($L$-band)
and $z=1.2$ ($M$-band),
where a deceleration parameter $q_0=0.1$ is used. The bottom 
panels are
similar to the middle ones, except that a deceleration parameter
of $q_0=0.5$ is adopted.
A remarkable similarity in the restframe K$'$ images of NGC 922 is 
found,
independent of redshift and of the deceleration parameter assumed.}
\label{fig1}
\end{figure}

\begin{figure}
\centerline{\includegraphics[width=35pc]{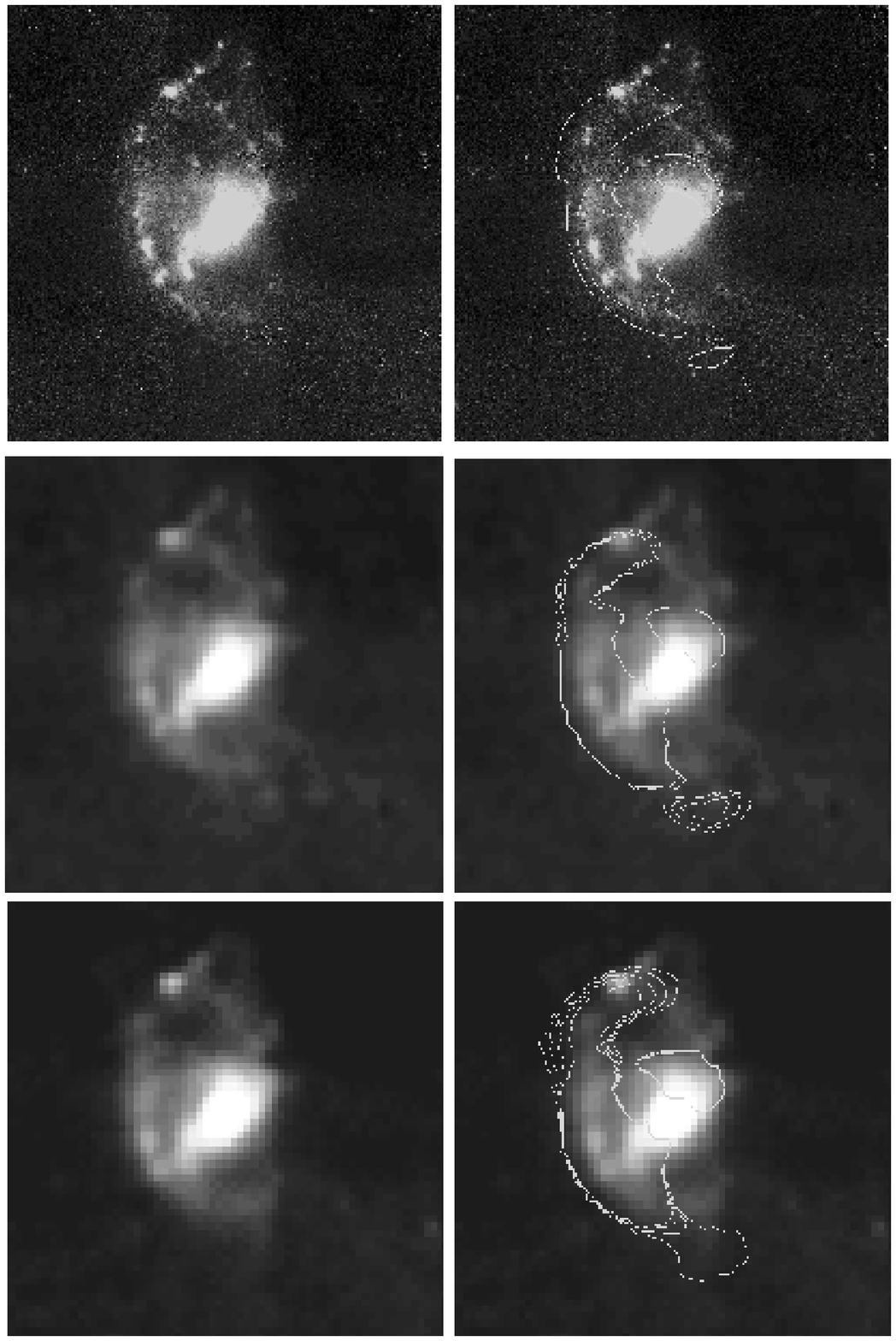}}
\vskip-60pt
\caption[]{Dust penetrated images of NGC 922. Upper left: 
Groundbased
K$'$ image. Middle left:
simulated 1h exposure (assuming an 8m NGST), at redshift 0.7 (L
band). Bottom
left: NGC 922 simulated at redshift z=1.2 (M band) with an 8m 
NGST. The
right
hand panels show contour overlays of the low order $m$=1 and 2 
inverse
Fourier transforms.} \label{fig2} \end{figure}

\section{Conclusions}

In this paper, we have simulated images of NGC 922 when
it is systematically moved
out in redshift space, always in restframe K$'$. Our
analysis confirms that the morphology of such optically peculiar 
objects
may successfully
be described by low-order $m$ modes, as for galaxies at $z\sim
0$ in our local Universe.

\end{article}
\end{document}